\documentclass{mn2e}
\def \th {\thinspace}

\def \degmark{^\circ}

\def\approxgt{\mathrel{\hbox{\rlap{\lower.55ex \hbox {$\sim$}}
\kern-.3em \raise.4ex \hbox{$>$}}}}
\def\approxlt{\mathrel{\hbox{\rlap{\lower.55ex \hbox {$\sim$}}
\kern-.3em \raise.4ex \hbox{$<$}}}}
\def \th {\thinspace }
\def \ref {\reference{}}

\def \sun {\hbox {$\odot$}}
\def \degmark{^\circ}

\usepackage{graphicx}

\title[Measurements of ADC radius]             
{Measurements of accretion disc corona size in LMXB:\\
consequences for Comptonization and LMXB models}

\author[Church and Ba\l uci\'nska-Church]    
{M. J. Church$^{1,2}$ and M. Ba\l uci\'nska-Church$^{1,2}$\\
      $^1$University of Birmingham, School of Physics and Astronomy,
      Birmingham, B15 2TT, UK\\
      $^2$Astronomical Observatory, Jagiellonian University, ul. Orla 171,
      30-244 Cracow, Poland}

\date{Accepted 2003 September 4. Received 2003 June 11}

\begin{document}
\maketitle


\begin{abstract}
We present results of measurements of the radial extent of the accretion
disc corona in low mass X-ray binaries, i.e. of the radial extent of the
thin, hot ADC above the accretion disc. These results prove conclusively 
the extended nature of the ADC, with radial extent varying from 20,000 km 
in the faintest sources to 700,000 km in the brightest, a substantial fraction 
of the accretion disc radius, typically 15 per cent. This result {\it rules
out the Eastern model for LMXB} which is extensively used, in which the 
Comptonizing region is a small central region. The ADC size depends strongly 
on the 1 -- 30 keV source luminosity {\it via} a simple relationship 
$r_{\rm ADC}$ = $L^{0.88\pm 0.16}$ at 99 per cent confidence, which is close to 
a simple dependence $r_{\rm ADC}$ $\propto$ $L$. We also present limited
evidence that the ADC size agrees with the Compton radius $r_{\rm C}$, or maximum radius
for hydrostatic equilibrium. Thus, the results are consistent with models
in which an extended ADC is formed by illumination of the disc by the central 
source. However, the dependence on luminosity may reflect the known {\it decrease}
of coronal temperature as the source luminosity increases leading to an
increase of $r_{\rm C}$. The extended nature of the ADC means that the seed 
photons for Comptonization {\it must consist} of emission from the disc to 
the same radial extent as the corona, providing copious supplies of soft 
seed photons. We thus demonstrate the importance of the size of the ADC to 
the correct description of Comptonization, and we derive the Comptonized 
spectrum of a LMXB based on the thermal Comptonization of these seed photons
and show that this differs fundamentally from that of the Eastern model, which 
assumes a cut-off in the spectrum below 1 keV.
Finally, we argue that our results are inconsistent with the assumption often made 
that the X-ray emission of accreting Black Holes and Neutron Stars has a {\it common mechanism
depending on the properties of the accretion flow only}. 
\end{abstract}


\section{Introduction}

A major problem that has impeded understanding of Low Mass X-ray Binaries 
containing neutron stars over many years
has been the controversy over the location and nature of the X-ray emission regions. Two radically
different descriptions were developed in the mid-1980s, the Western model and the Eastern
model. In the Western model, the dominance of Comptonization in the spectra of LMXB was
acknowledged by modelling the spectra with the Generalised Thermal model having the form
of a power law, cut off at high energies corresponding to the energy limit of Comptonizing
electrons (White, Stella \& Parmar 1988). 
The Eastern model assumes multi-temperature blackbody 
emission from the inner disc, plus emission from the neutron star providing the seed photons
for the observed Comptonized emission (Mitsuda et al. 1989).
Comptonization is assumed to take place either in the atmosphere of the neutron star 
or in a small, central region containing hot
electrons. Both Western and Eastern model were generally capable of
providing good fits to the spectra of LMXB, and so could not be discriminated between on this
basis, although the values of model parameters obtained are sometimes not
physically acceptable (see below). 
In the 1990s, there was increasing evidence that two emission components were present 
in all sources, whereas the Western model only needed a second component in bright sources.
A new model (the Birmingham model) closely related to the Western model was proposed 
in which two continuum components exist in all LMXB,
simple blackbody emission from the neutron star plus Comptonized emission from an
extended accretion disc corona (ADC) above the accretion disc (Church \& Ba\l uci\'nska-Church 1995).
The addition of the
blackbody results in markedly different spectral fitting results compared with the
Western model for parameters such as the cut-off energy 
(e.g. Church et al. 1998b).
Clearly, the physical descriptions contained in the Birmingham model and the Eastern model 
are radically different and the failure to resolve which model is correct has
been a major impediment to understanding. This includes understanding of the {\it phenomena}
in LMXB, for example, the physical changes taking place
during track movement in the Z-track and Atoll sources.

The Eastern model is extensively used.
For example, the model has been applied in analysis of {\it BeppoSAX} data of XB\th 1658-298
(Oosterbroek et al. 2001a), for GX\th 3+1 and Ser\th X-1 (Oosterbroek et al. 2001b),
and in studies of globular cluster sources (Sidoli et al. 2001).
Done, \.Zycki \& Smith (2002) argued for the model on theoretical grounds and then applied 
the Eastern model to analysis of data on 
Cyg\th X-2, as was the case for 4U\th 1608-52 (Gierlinski \& Done 2002).
Barret, Olive \& Oosterbroek (2003) used the Eastern model for {\it BeppoSAX} and {\it RXTE}
observations of 4U\th 1812-12, but concluded that the blackbody emission was
from the neutron star. Narita, Grindlay \& Barret (2001) carried out spectral fitting 
of {\it ASCA} observations of GX\th 354-0
and KS\th 1731-260 and give a detailed discussion in terms of the models above.
The physical description of the emitting regions can be made very specific by proponents of
the Eastern model. For example,
Done et al. (2002) assume that there is an intrinsic low energy cut-off in the
spectrum at $\sim$1 keV due to lack of low energy seed photons, and that the emission
of the neutron star is probably buried beneath an optically thick boundary layer,
leaving the disc and the boundary layer as the emitters. Seed photons can be
from the inner disc or neutron star surface. It is assumed that the disc dominates the spectrum
at low energies and the Comptonized boundary layer dominates at high energies.
These assumptions have
a major effect on the description of the X-ray emission, and will be reviewed in the
context of the results of the present work in Sect. 4.

Theoreticians have in general considered it natural that 
the Comptonizing region will be located close to the neutron star and so have supported 
the Eastern model (e.g. Kluzniak \& Wilson 1991; Popham \& Sunyaev 2001). 
Moreover, based on the similarity of the spectra of black hole binaries (BHB)
and neutron star binaries that is sometimes claimed, and 
on similarities in their timing properties, there
has been a tendency to assume a common mechanism for X-ray formation (e.g. Poutanen 2001).
To be independent of the type of compact object, the mechanism must
be a property of the accretion flow only. However, this makes the assumption that
the X-ray emission of the neutron star does not modify substantially the 
geometry, properties and X-ray emission of the ADC. In the present work, we show that
the evidence is against this assumption. Moreover, in neutron star systems
there is strong evidence for neutron star blackbody emission, while in BHB
there is disc blackbody emission. 

In fact, the key to resolving the controversy of the correct emission model has lain
with the dipping LMXB which provide more diagnostics of the emission regions
that non-dipping sources and more strongly constrain spectral models.
These sources having inclination angle between 
65$\degmark$ and 85$\degmark$ exhibit orbital-related X-ray dips due to absorption 
in the bulge in the outer disc (White \& Swank 1982; Walter et al. 1982).
An acceptable description of 
spectral evolution in dipping in these sources requires fitting not only the non-dip spectrum 
but also the spectra of several levels of dipping, selected typically in intensity bands.
Moreover, only absorption parameters can be allowed to vary in the fitting, and emission
parameters must be held constant, so that fitting strongly constrains
emission models in these sources. On the basis of such work, the Birmingham model
was proposed by Church \& Ba\l uci\'nska-Church (1993, 1995).
It was apparent from the start that one emission component was very extended, 
since this non-thermal component
of the spectra was removed very slowly in dipping as the extended absorber overlapped an
extended emission region (e.g. Church et al. 1997).
It also became apparent that the ADC was {\it thin} (i.e. $H/r$ $<<$ 1) since it would
be very unlikely that 100 per cent deep dipping would be observed if the extended ADC was
spherical, this requiring an absorber on the outer disc extending out of the orbital 
plane to very large distances.
Application of the new model provided substantial
evidence that the blackbody component was present in {\it all} sources in
varying degrees. Over a period of years it has been shown that the model provides very good 
fits to all of the $\sim$10 dipping sources (e.g. Church et al. (1997, 1998a, 1998b); 
Ba\l uci\'nska-Church et al. (1999, 2000); Smale, Church \& Ba\l uci\'nska-Church (2001, 2002);
Barnard, Church \& Ba\l uci\'nska-Church (2001). It also
fits well the Atoll and Z-track sources included in an {\it ASCA} survey of LMXB
(Church \& Ba\l uci\'nska-Church 2001) showing that the model describes well all 
classes of LMXB.
Thus both the Eastern and the Birmingham models are two-component models, each having
a thermal and a Comptonized component. However, it is difficult
to decide from spectral fitting (a single spectrum) between the models.
The crucial factor is, in fact, the origin of the Comptonized emission, which
in the Eastern model is a small central region, i.e. of the order 
\begin{figure*}                                             
\begin{center}
\includegraphics[width=78mm,height=150mm,angle=270]{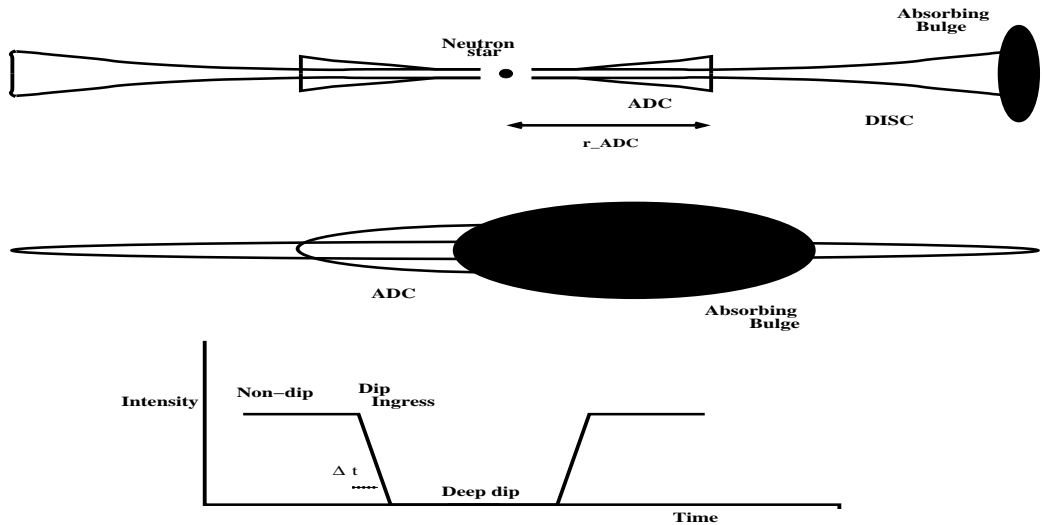}    
\caption{Geometry of a LMXB showing the extended accretion disc corona of radial extent $r_{\rm ADC}$
above a thin accretion disc, and the bulge in the outer disc where the accretion flow from the
companion star impacts, which is responsible for X-ray dipping. The central diagram shows 
the view at a large inclination angle ($\sim$75$\degmark$) typical of dipping sources
as seen by an observer of a bulge of larger angular size than the ADC which is partly
covering the ADC, and will lead to 100 per cent deep dipping takes place, with ingress
time $\Delta$t as in the lower diagram.}
\end{center}
\end{figure*}
of 100 km 
radius or less, and in the model of Church and Ba\l uci\'nska-Church
is very extended of radial extent typically 50000 km (Church 2001), i.e. 500 times
larger. In the dipping sources, measurement of emission region sizes by the
technique of dip ingress timing is possible, and this clearly holds the
key to discriminating between these two models. Under conditions where the
angular size of the absorber is larger than that of the extended emitter,
the ingress time is determined by the size of the source. This condition will hold
if dipping is 100 per cent deep at any energy, and we have applied the technique
to several of the dipping LMXB (Church et al. 1998a, 1998b; Ba\l uci\'nska-Church            .
et al. 1999, 2000; Smale et al. 2001, 2002).
In the present work, we have assembled our previous measurements of ADC size
and have analysed further observations providing good quality dip data, 
and present the results in Sect. 2. These results do not allow
the Eastern model to be possible. In the second part of the
paper, we deal with the consequences of this. Specifically, the
{\it correct Comptonization model for LMXB or BHB depends strongly on the
size of the Comptonizing ADC}. We compare
the Comptonized emission for the Eastern model with a small
central Comptonizing region, with that taking place in the actual extended
ADC revealed by our measurements. We thus show that the description of
Comptonization in the Eastern model is not correct, and that use of this
model in analysis will not produce correct results.

\section{Observations and analysis}

The observations selected consist primarily of data obtained using the satellites 
{\it ASCA}, {\it Rosat}, {\it BeppoSAX} and {\it Rossi-XTE} previously analysed 
by us in detail as shown in Table 1.
Apart from our proprietary observations, data are included resulting from
our collaborations with {\it ISAS}, Japan, {\it ESTEC} in The Netherlands 
and {\it LHEA} in the USA. High quality data were used as discussed below, and 
we note that data from {\it Chandra} or {\it XMM} will not necessarily provide
better values of ingress times, which vary between 100 and 16000 seconds, i.e. are
not difficult to measure. Figure 1 shows a schematic of a LMXB (not to scale)
including the
accretion disc, the corona above it covering the inner 15 per cent of the disc
and the bulge on the outside of the disc responsible for dipping.
The technique of dip ingress timing determines the overall radial extent $r_{\rm ADC}$ of
the corona, i.e. of the continuum emitting region. It cannot determine whether
the ADC has any ``inner'' radius. The only way in which ingress times
might not provide true values for the radial extent of the corona would be if
the absorbing bulge on the outer accretion disc had a complex structure resulting
in the gradual removal of intensity from a point-like corona. Attempts to model
dipping in XB\th 1916-053 on this basis failed to reproduce the observed light
curves (\.Zycki, private communication, 2001) and it appears that this possibility
can be discounted.
Also shown is a schematic of dip ingress and egress with ingress time $\Delta$t
for the case that the angular size of the absorber is larger than that of the
extended emitter, the ADC, so that dipping is 100 per cent deep.
This was the main criterion for inclusion of data, that
dipping had to be $\sim$100 per cent deep in any energy band covered by the instrument used.
Weak dipping at the 10 per cent level cannot be used as we cannot be certain
that the absorber angular size is greater than the ADC angular size. Thus, 
data on the sources X\th 1755-338 and XB\th 1746-371 was not used,
although data were available from {\it Exosat} and {\it ASCA}.
Similarly, {\it Rossi-XTE} data or XB\th 1323-619
had only 50 per cent deep dipping. 

We also rejected
data in which dip ingress was poorly defined, not allowing sensible measurement
of ingress times, such as the {\it Rosat} observation of XBT\th 0748-676. In
the observations included in Table 1, typically containing 4 or more
dips per observation, ingress times were only included in analysis for which
the ingress was well-defined. Frequently this was not the case if
Earth occultation or SAA passage caused data gaps during dipping. 
In observations containing several 
well-defined dips, such as the {\it BeppoSAX} observation of XB\th 1916-0563, 
an error estimate was obtained from the scatter of $\Delta$t about the mean.
In XBT\th 0748-676, we have revised upwards our previous value for $\Delta$t
(Church et al. 1998a).

The radius of the ADC, $r_{\rm ADC}$ is related to the accretion disc radius
$r_{\rm AD}$ {\it via} the equation:

\begin{equation}
{2\pi \,r_{\rm AD}\over P} =  {2\, r_{ADC}\over \Delta t}
\end{equation}

\noindent
where $P$ is the orbital period. In each source we used the latest values of
orbital period, including a period of 3000.6508 s for XB\th 1916-053 (Chou, Grindlay \& Bloser 2001),
and 20.8778$\pm$0.0003 hr for X\th 1624-490 from our recent work on this source
(Smale et al. 2001).
The radius of the sphere having the same volume as the Roche lobe of the neutron star
$r_{\rm L1}$ was calculated from 
the expression of Eggleton (1983)
accurate to 1 per cent:

\begin{equation}
{r_{\rm L1}} = {0.49\,a\,(M_1/M_2)^{2/3}\over{0.6\,(M_1/M_2)^{2/3} + ln(1+(M_1/M_2)^{1/3})}}
\end{equation}

\noindent
where $a$ is the separation of the stars in the binary system, 
%
\begin{table*}
\begin{center}
\begin{minipage}{180mm}
\caption{Values of Accretion Disc Corona radius derived from measurements of dip ingress times, together with
source luminosities and the parameters of the binary systems used. The fraction $f$ is the ratio
$r_{\rm ADC}/r_{\rm AD}$.}
\begin{tabular}{lllrrrrrrcr}
\hline
\hfil source&satellite&date&$\rm {L^{1-30}_{\rm Tot}}$&$d$&
$\rm {\Delta t}$&$P$&$r_{\rm AD}$&$r_{\rm ADC}$&$f$&ref.\\
&&&erg s$^{-1}$&kpc&s&hr&10$^{10}\,$cm&10$^9\,$cm&\%&\\
\hline
XB\th 1916-053  &Asca    &1993, May 2  &$4.11\times 10^{36}$&9.0&160$\pm48$   &0.834& 2.02 & 3.39 &16.8&1\\
XB\th 1916-053  &Rosat   &1992, Oct 17 &$3.28\times 10^{36}$&9.0&112$\pm34$   &0.834& 2.02 & 2.37 &11.7&2\\
XB\th 1916-053  &SAX     &1995, Oct 13 &$3.91\times 10^{36}$&9.0&146$\pm15$  &0.834& 2.02 & 3.09 &15.3&3\\
XBT\th 0748-676 &Asca    &1993, May 7  &$3.98\times 10^{36}$&10.0&280$\pm84$  &3.820& 4.07 &2.60 &6.4 &4\\
XB\th 1323-619  &SAX     &1997, Aug 22 &$3.21\times 10^{36}$&10.0&254$\pm57$  &2.938& 3.51 &2.65 &7.5 &5\\
XB\th 1254-690  &RXTE    &2001, May 9  &$2.18\times 10^{37}$&12.0&950$\pm190$  &3.933& 4.14 & 8.72 &21.1&6\\
XB\th 1254-690  &Exosat  &1984, Aug 6  &$1.25\times 10^{37}$&12.0&840$\pm250$   &3.933& 4.14 &7.71 &18.6&7\\
X\th 1624-490   &RXTE    &1999, Sep 27 &$1.44\times 10^{38}$&15.0&12500$\pm2500$&20.98& 11.15 &58.0 &52.2&8,9\\
X\th 1624-490   &Exosat  &1985, Mar 25 &$1.10\times 10^{38}$&15.0&15500$\pm2500$&20.98& 11.15 &71.9 &64.8&10\\
\hline
\end{tabular}
\end{minipage}
\end{center}
\hfil\hspace{\fill}
References: $^1$ Church et al. 1997; $^2$ Morley et al. 1999;
$^3$ Church et al. 1998a; $^4$ Church et al. 1998b;  $^5$ Ba\l uci\'nska-Church et al. 1999;\\ 
$^6$ Smale et al. 2002; $^7$ present work; $^8$ Ba\l uci\'nska-Church et al. 2000;
$^9$ Smale et al. 2001; $^{10}$ Church \& Ba\l uci\'nska-Church 1995.
\end{table*}
\begin{figure*}                                             
\begin{center}
\includegraphics[width=78mm,height=150mm,angle=270]{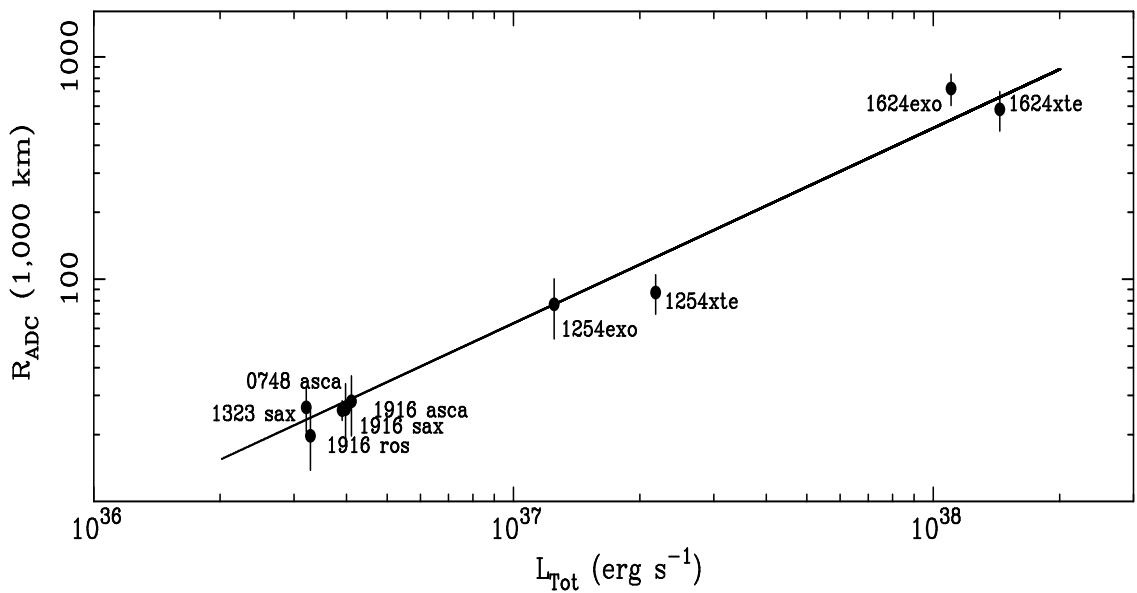}
\caption{Measured values of ADC radius as a function of total source luminosity in the band 1 -- 30 keV.}
\end{center}
\end{figure*}
%
$M_1$ is the mass of
the neutron star, and $M_2$ is the mass of the companion star.

Masses of the companion stars were found using appropriate mass-period relationships,
using the form for
Main Sequence stars in the case of XB\th 1323-619, XB\th 1254-690, X\th 1624-490 
and XBT\th 0748-676. 
It was assumed that the uncertainty in the mass was about 10 per cent; however, tests
showed that the corresponding error in
$r_{\rm L1}$ was reduced to 1 per cent.  The accretion disc radius $r_{\rm AD}$ was taken to be
$0.8\,r_{\rm L1}$ based on values $0.74\,r_{\rm L1}$ -- $0.84\,r_{\rm L1}$ (Armitage \& Livio 1996)
and $0.9\,r_{\rm L1}$ (Frank et al. 2002). In the case of XB\th 1916-053,  
the companion star is likely to be a hydrogen-deficient star, 
neither He-burning nor fully degenerate (Nelson, Rappaport \& Joss 1986), with mass 0.10 -- 0.15
M$_{\sun}$, and we calculated $r_{{\rm L1}}$ for 0.125 M$_{\sun}$.
This is insensitive to the mass of the companion so that if it were assumed that 
the companion was a low mass white dwarf as suggested by White \& Swank (1982),
the change would be only 10 -- 25 per cent depending on composition.
The main source of uncertainty in the ADC radial extent obtained is the
error in the ingress times, varying between 10 and 30 per cent.
Where possible, the error was obtained from the scatter of values. For observations providing
only a single $\Delta$t, the error was assumed to be 30 per cent. Results are shown 
in Table 1.

Spectral fitting results previously obtained by us were used to provide the source fluxes
in the band 1 -- 30 keV, chosen to include as much of the spectrum as possible without requiring
large extrapolations for data limited to 10 keV. Total source luminosities 
$L^{1-30}_{\rm Tot}$ in this band are shown in Table 1. 
In the case of the {\it Exosat} observation of XB\th 1254-690, we obtained
a background-subtracted spectrum from the {\it HEASARC} archive, and fitted this
with the Birmingham two-component model, as the original fitting had used a one-component
model. For the {\it Exosat} observation of X\th 1624-490,
spectral fitting the original data would be limited by the 1 -- 10 keV range of the ME.
Because of this, we adopted our solution for the {\it RXTE} observation (Smale et al. 2001) having
similar luminosity, and scaled the 1 -- 30 keV luminosity by the ratio of the
predicted and observed count rates in {\it Exosat}. Using recent values of source distances, mostly
obtained from Christian \& Swank (1997), fluxes were converted to 1 -- 30 keV luminosities.
In Fig. 2, we show the variation of ADC radial extent with $L_{\rm Tot}$.

\section{Results}

Fig. 2 displays two significant results: it confirms that the ADC is very extended
as is also apparent from spectral fitting of dip spectra (Sect. 1), and secondly, there
is a strong correlation between $r_{\rm ADC}$ and source luminosity, although this
may be a direct dependence due to X-ray irradiation, or an indirect dependence on $\dot M$.
The extended 
ADC varies between a minimum of $\sim$7 per cent of the accretion disc radius to a maximum 
of 65 per cent (in X\th 1624-490).
In the case of XB\th 1916-053, our measured ingress times fall within the range
of values obtained by Narita et al. (2003), who also found some evidence for a
dependence of ADC size on mass accretion rate in data on this one source
having limited variation in luminosity.
Least squares fitting provides the power law dependence $r_{\rm ADC}$
= $L_{\rm Tot}^{0.88\pm 0.16}$ at 99 per cent confidence, suggesting a simple proportionality.

We wish, of course, to understand the formation of the ADC, which previously has been very
unclear, or controversial, and 
this will be discussed in Sect. 5. However, one issue will be whether the increase
is size with luminosity is due to increasing illumination of the disc, or whether
the ADC size is limited by the fact that hydrostatic equilibrium is not possible
outside a limiting radius.
For a small number of sources, we have broadband spectra obtained from {\it BeppoSAX}
extending to 100 keV which allowed the mean Comptonization cut-off energy to be obtained,
and from this the mean electron temperature $T_{\rm e}$ of the Comptonizing ADC. Information on the
ADC from observation is, of course, quite limited, and we can do no more than obtain
mean quantities. However, theoretically, a strong radial variation of 
$T_{\rm e}$ is not expected. In a corona with $kT$ $\sim$10 keV, it is expected that
at radius $r_{\rm C}$, the Compton radius, hydrostatic equilibrium will fail, since

\begin{equation}
{kT} > {GMm_p/r \;\;\;\;\;\;\;   {\rm and} \;\;\;\;\; r_{\rm C} \sim GMm_p/kT  }
\end{equation}

\noindent
where $m_{\it p}$ is the proton mass, and outside this radius, the ADC will
dissipate as a wind. In reality, hydrostatic equilibrium may begin to fail
at somewhat smaller radial distances (e.g. Woods et al. 1996).
In Table 2, we show values of $r_{\rm C}$
calculated for the two cases: a high optical depth ($\tau$) corona, in which 
the Comptonization cut-off energy $E_{\rm CO}$
= 3 $kT_{\rm e}$, and a low optical depth corona, for which $E_{\rm CO}$ = $kT_{\rm e}$.
It can be seen that the values are substantially less than the radii of the accretion
disc in these sources, the high-$\tau$ values being about 30 per cent of the disc radius
and the low-$\tau$ values being about 10 per cent. 
\tabcolsep 1.5 mm
\begin{table}
\begin{center}
\begin{minipage}{80mm}
\caption{Approximate values of the Compton radius or limiting radial size of an ADC
in hydrostatic equilibrium for sources having well-determined $E_{\rm CO}$.
Ranges of electron temperature are given, with the lower limit for high optical
depth in the corona, and the higher limit for low optical depth. Corresponding values of
$r_{\rm C}$ are given in the next two columns and can be 
compared with approximate values of $r_{\rm ADC}$ (last column) from Table 1.}
\begin{tabular}{lrrrrr}
\hline
\
 & \hfil $\rm {E_{CO}}$ & $kT_{\rm e}$ & $r_{\rm C}$ & $r_{\rm C}$ & $r_{\rm ADC}$\\
 &&& High $\tau$ & Low $\tau$\\
 & keV & keV & 10$^9$ cm & 10$^9$ cm & 10$^9$ cm\\
\noalign{\smallskip\hrule\smallskip}
XB\th 1916-053  &80 & 26.7 - 80.0 & 7.2   &  2.4 & 3.0  \\
XB\th 1323-619  &44 & 14.7 - 44.0 & 13.1  &  4.4 & 2.7  \\
X\th 1624-490   &12 &  4.0 - 12.0 & 48.2  & 16.1 & 60 \\
\noalign{\smallskip}
\hline
\end{tabular}
\end{minipage}
\end{center}
\hfil\hspace{\fill}
References: Church et al. 1997; Ba\l uci\'nska-Church et al. 1999, 2000.
\end{table}

It can be seen that for the bright source X\th 1624-490, there is good agreement
between the measured size of the ADC (Table 1) and the high-$\tau$ value of
the limiting radius. For the other two, relatively faint, sources XB\th 1916-053
and XB\th 1323-619, the agreement is better with the low-$\tau$ value.
It is not appropriate to give a detailed discussion based on 3 sources; however,
there is clearly some evidence that the ADC size is limited to the maximum hydrostatic
size. In the case of X\th 1624-490, the large value of $r_{\rm C}$ is
a consequence of the {\it low} value of coronal electron temperature. 
Formation and size of the ADC is further discussed in Sect. 5.1; however
the large values of $r_{\rm ADC}$ measured are consistent with theoretical tretments
in which the ADC is formed by illumination of the disc by the central X-ray source.

There are two major consequences of the results: firstly the extended ADC size does not
allow the possibility of the Eastern model in which Comptonization is assumed to take
place in a central region associated with the neutron star or inner disc. The radius of
such a region would be of order $\approxlt$100 km, which is 500 times smaller than
measured here. Secondly, it will be seen in the next section that 
Comptonization is strongly dependent on the size of the ADC, and
the very soft nature of the seed photons generated in the disc below the extended ADC 
will be seen, contrasting with the 1 keV seed photons assumed in the Eastern model.

\begin{figure*}
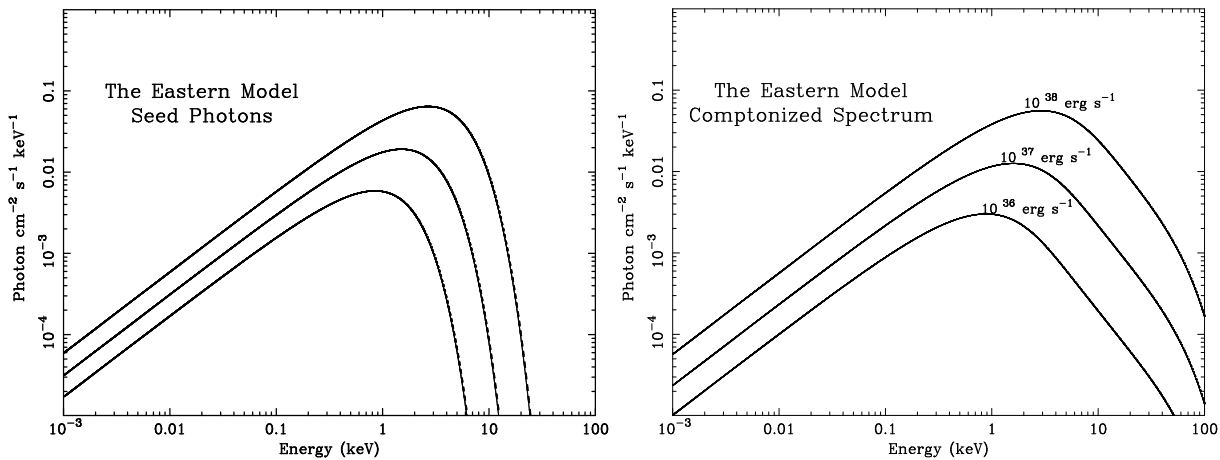
                                             
\includegraphics[width=60mm,height=80mm,angle=270]{f3a}     
\includegraphics[width=60mm,height=80mm,angle=270]{f3b}     
\caption{The Eastern model. The seed photons consist of simple blackbody emission originating
on the neutron star or inner accretion disc as used by proponents of the Eastern model.
The Comptonized spectra corresponding to the seed photon spectral forms
are shown for three luminosities in the band 1 -- 30 keV (see text).}
\end{figure*}
\begin{figure*}
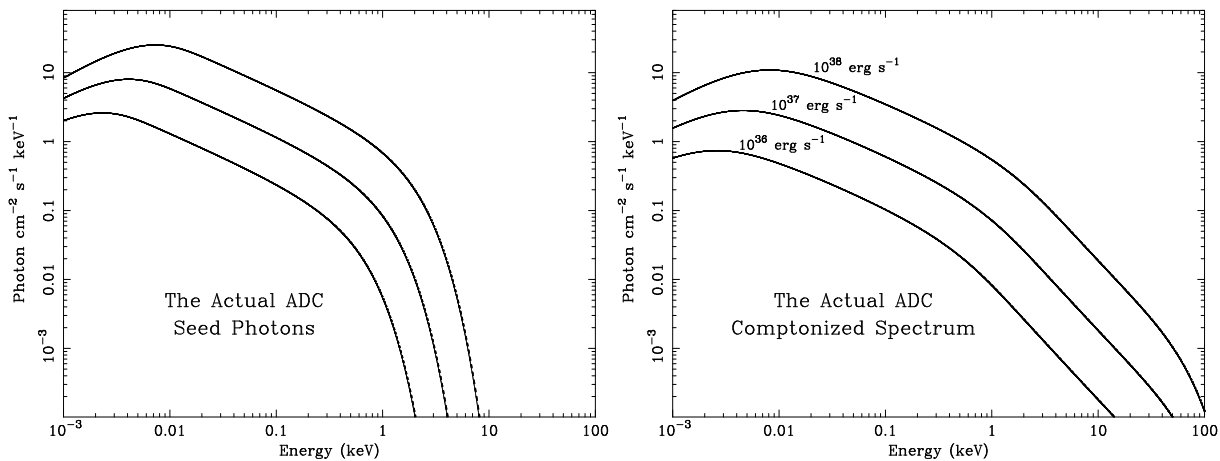
                                              
\includegraphics[width=60mm,height=80mm,angle=270]{f4a}      
\includegraphics[width=60mm,height=80mm,angle=270]{f4b}      
\caption{The actual ADC with size as measured  in the present work. The seed photon spectrum
(left panel) is very soft as emitted by an accretion disc between 10 and 50000 km.
The Comptonized emission (right panel) corresponds to the three 1 -- 30 keV luminosities
indicated.}
\end{figure*}
\section{The dependence of Comptonization on ADC size}

\subsection{Seed photons and Comptonized spectra}

In the following, we compare the Eastern model with the Birmingham model, i.e. a model
having an extended ADC as revealed by the present work.
In Figs. 3 and 4, we show the seed photon spectra (left panels) and the corresponding
Comptonized spectra (right panels). The Comptonized spectra are shown for
1 -- 30 keV luminosities of $10^{36}$, $10^{37}$ and 
$10^{38}$ erg s$^{-1}$, for a source at 10 kpc.
To allow direct comparison, we keep the ADC size
constant at 50000 km; however, allowing this to increase would only 
affect the spectrum below 0.01 keV.

In the Eastern model,
it is assumed that seed photons originate on the neutron star or inner disc
and so are modelled by a simple blackbody with $kT$ $\sim$1 keV (e.g. Done et al. 2002).
Typical values of $kT$ of 0.53, 0.94 and 1.67 keV were chosen which would produce
luminosities of $10^{36}$, $10^{37}$ and $10^{38}$ erg s$^{-1}$ assuming emission
from the whole neutron star. For convenience
we obtain the approximate form of the
Comptonized spectrum for this seed photon input using
the {\sc thcompfe} thermal Comptonization model (Zdziarski, Johnson \& Magdziarz 1996)
which is a non-standard model in {\it Xspec} which assumes a simple blackbody
for the seed photons. 
An electron temperature $kT_{\rm e}$ of 20 keV was used, representative of
values derived from broadband spectra between 5 and  100 keV, and a typical power law photon 
index of 1.7. The Comptonized spectrum of the Eastern model 
is shown in Fig. 3 (right panel) with normalizations in the model adjusted to
provide the three standard luminosities above. The normalization of each seed photon
spectrum was chosen such that the bolometric blackbody luminosity
was one fifth of that of the Comptonized spectrum, i.e. adopting a reasonable value for the
amplification factor.

For the extended ADC, we calculated the multi-temperature
disc blackbody emission of the seed photons by integrating to $r$ = 
50000 km. This was done assuming the temperature profile of a Shakura-Sunyaev
thin disc (Shakura \& Sunyaev 1976), making the
zero torque assumption that has the effect of causing $T(r)$ to decrease at the 
inner disc (Abramowicz \& Kato 1989). We again assume a
value for the amplification factor, but this is not critical as we only wish
to show the relation between the forms of the seed and output spectra.
We assumed that the total emitted X-ray luminosity
of the Comptonized emission
$L_{\rm tot}$ = $L_{\rm s}$ + $L_{\rm h}$, where $L_{\rm s}$ is the luminosity of the seed
photons and $L_{\rm h}$ is the heating supplied by the electrons, giving an amplification
factor $A$ = $L_{\rm tot}/L_{\rm s}$, and we assumed $A$ = 5. Thus
the bolometric seed photon luminosity was made one fifth of that of the Comptonized
emission, and
from the equivalent mass accretion rate, the temperature distribution in  the
disc was calculated. A non-standard {\it Xspec} model was produced
allowing the disc blackbody emission with this temperature profile
to be integrated between 10 km, the surface of the neutron star, and 50000 km.  

To obtain the Comptonized spectrum of these seed photons,
we made a dedicated {\it Xspec} model based on the thermal Comptonization model
{\sc thcompds} (derived from {\sc thcompfe} and provided by P. \.Zycki)
which calculates the Comptonized spectrum 
for seed photons produced by the complete accretion disc, 
assuming the {\sc diskbb} form
for the seed photons (Mitsuda et al. 1989) in which the zero torque assumption is
{\it not} embodied. Our modification involved changing the input seed photon spectrum
to correspond to the inner 15 per cent of the disc, and we produced two versions: with and
without the zero torque assumption. Basing our model on
{\sc thcompds} was done for convenience to show
the approximate form of the Comptonized spectrum, but assumes low optical depth,
whereas our evidence is for high optical depth in the ADC (Church 2001). However, this will not
affect the major conclusions of the present work. An electron temperature
$kT_{\rm e}$ of 20 keV was assumed, as before.
In Fig. 4 we show the spectra of the seed photons
and the Comptonized spectrum assuming zero torque for the extended ADC.

Comparison of Figs. 3 and 4 reveals the strong differences between 
Comptonization in the Eastern
model and in the Birmingham model with an extended ADC; for clarity we do not include
the second thermal component also present in each model. 
The difference between the seed photon spectra
are extreme, in the Eastern model consisting of a blackbody of $kT$ $\sim$1 keV,
but for the actual extended ADC consisting of a disc blackbody peaking at 0.003 keV for
the lowest luminosity and at 0.01 keV for the highest luminosity.
These differences have a major effect on the Comptonized spectra such
that in the Eastern model, the spectrum {\it decreases} below a peak at 1 -- 3 keV
depending on luminosity, reflecting a widely-held view that a Comptonized
X-ray spectrum cannot continue to follow a power law at low energies
because of an expected shortage of seed photons. However, the consequence of an 
extended ADC is that the Comptonized spectrum below 1 keV {\it does} continue to rise
causing the Comptonized spectrum also to rise. Eventually there is a 
turn-over in the spectra at 0.01 keV, however, this is considerably
below the minimum energy in X-ray telescopes. A much less pronounced change in slope
takes place in the Comptonized spectrum at energies of $\sim$1 keV due to the
corresponding break in the seed photon spectrum. This change of slope
would be difficult to detect even with data extending to 0.1 keV because of
low energy absorption. Finally, the high energy cut-off in Fig. 4 can be seen
at $\sim$80 keV corresponding to the maximum energy of Comptonizing electrons.

\begin{figure}                                              

\includegraphics[width=60mm,height=80mm,angle=270]{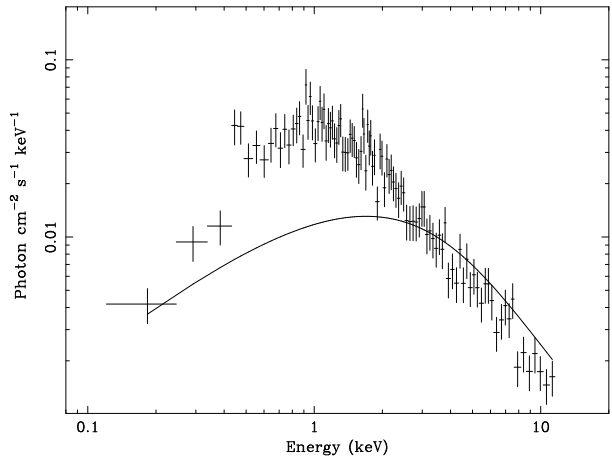}    
\caption{{\it XMM} EPIC data simulated assuming seed photons from the inner 15 per
cent of the accretion disk, fitted by a thermal Comptonization model appropriate to the 
Eastern model {\sc tthcompfe}. $kT$ for the seed photons was frozen at 1 keV (see text).}
\end{figure}

The zero torque assumption is appropriate to a black hole
binary where the accretion flow cannot interact with the stellar surface;
however, it is not clear whether it is appropriate in a neutron star system.
We have tested the effect of this assumption, and find it does not
affect our main results.
Not making the zero torque assumption is to modify the seed photon spectrum
by moving the energy of the slight break in slope from 1 keV to 1.5 keV
(for the central curve).


\subsection{Fitting with the Eastern model}

In the above, we compare the Comptonized spectrum expected for an extended ADC
with that of the Eastern model.
We now address the question of what effect this will have in spectral fitting
with the Eastern model. To answer this, we have simulated 20 ks of {\it XMM} EPIC data using the 
Comptonized spectrum of an extended ADC (as in Fig. 4, right panel) with parameter
values: $kT_{\rm e}$ = 20 keV, power law index = 1.7, peak disc temperature 0.4 keV,
column density $0.2\times 10^{22}$ atom cm$^{-2}$, an appropriate instrument response function,
and fitted this using the thermal Comptonization model corresponding to the Eastern model:
{\sc thcompfe}. We make this comparison of Comptonized spectra without adding further
emission components that exist in real data to the simulation.
It was arranged that $L^{1-30}$ was 10$^{37}$ erg s$^{-1}$.

In Fig. 5, we show the result of fitting this simulated data 
with the {\sc thcompfe} model.
Initially, we fixed the blackbody temperature at 1 keV as is typical
for results from using the Eastern model, and implies that seed photons can originate
on the neutron star, and so can have higher $kT$ than the inner disc temperature used
above. The model fits very badly, particularly below 2 keV
with $\chi^2$/d.o.f. = 628/100 as shown in Fig. 5. With $kT$ free, an acceptable fit was obtained, but with
a value of $kT$ = 0.14 keV, which is inconsistent with the seed photons assumed in the Eastern model.
Thus, it was not possible to fit the data sensibly with the {\sc thcompfe} model, as might be
expected, since the data were simulated assuming an extended disc blackbody source of
seed photons, and fitted by a localized source. The only way to obtain an acceptable fit
with $kT$ $\sim$1 keV was found to be by adding a second, unreal, model component (in fact, an extra
blackbody component) which was not present in simulating the data. 
We conclude that fitting the Eastern model is likely to produce incorrect results.
Fitting real data will be complicated by the presence of a second emission component,
which we associate with the neutron star.



\section{Discussion}

The present work provides strong evidence for the Birmingham model of LMXB,
and against the Eastern model. In particular, we have shown that
the Comptonizing region {\it cannot} be a small, central region, as in the 
Eastern model, but is, in fact, an extended, flat, hot region above the inner 15 per cent 
typically of the accretion disc, i.e. constituting a thin ADC above a thin disc.
We have also shown that the size of the Comptonizing region has strong
implications for the correct form of the Comptonization model to be used.
The relevance to ADC formation is discussed below; however, we first address briefly
two issues. The identification of blackbody emission in LMXB has been a problem:
the Eastern model identifies this with disc emission whereas we, and others,
identify this with emission from the neutron star. Given the large ADC, 
we expect all of the X-ray emitting inner disc to be covered by ADC. Moreover,
there is evidence for the high optical depth of the ADC plasma (e.g. Church 2001;
Narita et al. 2001), and it is likely that essentially all of the disc emission 
will be reprocessed in the ADC and so not observed, and that the
observed thermal emission originates on the neutron star. Secondly, several authors 
have proposed that X-ray emission in accreting black holes
and neutron stars has a common origin depending on the properties of the accretion
flow only, effectively a unifying model. For example, spectral similarities between low state 
black holes and neutron star binaries, and
similarities in timing properties suggest a common mechanism for X-ray
production (Poutanen 2001). In BHB, the thermal emission component
is clearly from the disc, whereas in LMXB, disc emission will be reprocessed in the ADC
and so not seen. Thus the present work does not support a common emission
mechanism, since this relies on the
assumption that disc blackbody emission is observed from each type.

\subsection{Formation of the ADC}

The dependence of ADC size on X-ray luminosity shown in Fig. 2 has implications for
understanding ADC formation. Various mechanisms for ADC production have been proposed.
Liang \& Price (1977) suggested that internal processes within the disc would transfer
energy to a corona with low density which would not radiate and cool efficiently. 
Paczy\'nski (1978) invoked gravitational instability in the disc.
Galeev, Rosner \& Vaiana (1979) proposed corona formation by expulsion of magnetic
field loops from the disk; see also Beloborodov (1999) and Miller \& Stone (2000).  
In addition, there is the two-phase model of a hot corona above a cool disc of 
Haardt \& Maraschi (1993), and the hot disc model (Shapiro, Lightman \& Eardley 1976;
Zdziarski 1998).
Shakura \& Sunyaev (1973) suggested an external process in which the hot, inner disc
would evaporate disc into corona. White et al. (1981) suggested that in the ADC
source 4U\th 1822-371, X-rays from the central source were scattered in a corona
evaporated by X-rays from the disc (see also White \& Holt 1982). Fabian, 
Guilbert \& Ross (1982) calculated the
density and optical depth of an atmosphere above the disc. Begelman, McKee \& Shields (1983)
presented detailed calculations of the formation of the corona by irradiation
from the central source (see also Begelman \& McKee 1983).
Recently, there has been further detailed modelling
of this kind. R\'o\.za\'nska \& Czerny (1996) and R\'o\.za\'nska et al. (1999)
investigated the effects of illumination by the central source on the accretion disc 
for either a black hole as in AGN or a neutron star as in Galactic binary sources.
It was found that the disc and ADC were separated by a boundary layer
or heated upper region of the disc; i.e. the region above the disc became stratified 
with increasing temperature and decreasing density (see also R\'o\.za\'nska et al. 2002).
Jimenez-Garate, Raymond \& Liedahl (2002) also
calculated the vertical and radial structure of the disc, and derived the line
emission for comparison with high spectral resolution data from {\it Chandra} and {\it XMM}.
The modelling of R\'o\.za\'nska,
Czerny and co-workers concentrated on the vertical structure at a given radial
position and did not consider the radial extent of the ADC. Similarly, Jimenez-Garate
et al. calculated the vertical structure, but allowed the ADC to extend to the
same radius as the accretion disc.
Proga \& Kallman (2002) also addressed the effects of X-ray irradiation of the disk,
with the aim of explaining the UV emission of LMXB, not presently understood.
They concentrated on  the reprocessing of X-rays in the disc so as to boost the
UV emission of the disc and the possible driving of a disc wind by UV lines.
Then the UV-driven wind could be a source of the observed UV radiation. Ionization
by X-rays was found to prevent this mechanism, although winds could be produced
by thermal expansion and other effects. The part of
the disk considered was that between $1\times 10^9$ and $1\times 10^{10}$ cm,
the region typically between 0.03 and 0.3 of the accretion
disc radius. Proga \& Kallman considered that the Compton radius where a static
corona becomes an unbound wind would be $\sim$ $10^{11}$ cm, greater than the
disc size, equivalent to assuming a coronal temperature of $1\times 10^7$ K.
However, our work has shown that the electron temperature $kT_{\rm e}$ is
$\sim$ 5 -- 30 keV, or more, giving a Compton radius typically $\sim$ 2 --
10$\times 10^9$ cm, and a reasonable explanation of our measured ADC radial
extents is that they are determined by the Compton radius. Thus, we expect
that a wind will exist anyway over much of that part of the disc considered by 
Proga \& Kallman, but for a different reason: because the corona is unbound.

The present work clearly supports the modelling in which a very extended ADC is
formed by illumination of the disc by the central source. 
The observational results indicate a strong
dependence on $L_{\rm Tot}$, and there is some evidence (Table 2) that the
size is limited to the maximum radius for hydrostatic equilibrium, suggesting that
the modelling should also include this effect. If this effect is dominant,
the size varies inversely with coronal temperature, and it is because brighter
sources have lower Comptonization cut-off energies (e.g. Church et al. 1997;
Ba\l uci\'nska-Church et al. 2000) and so lower temperatures
that the ADC becomes much larger.

In terms of earlier proposals for ADC formation, such as an internal mechanism 
in the disc, the process would have to form an ADC of size varying substantially 
with mass accretion rate (and thus $L_{\rm Tot}$), and also
predict an extended ADC. In general, the processes proposed do not predict an extended
ADC. For example, the magnetic loop process (Galeev et al. 1979) is concentrated
in the inner disc. One difficulty in making this comparison is that the theoretical
treatments are often `local', i.e. do not consider the radial dependence. However,
it appears that internal processes would not produce an ADC of the radial extent
found in the present work. Thus we conclude that ADC formation by direct X-ray irradiation
is more likely, in which case both the inner disc and the neutron star
probably contribute to ADC formation.

\subsection{The correct form of the Comptonization term in spectral fitting}

We have shown that the present results imply a particular
form for the Comptonized emission in LMXB. 
In our previous work on LMXB, the Birmingham model
with spectral form {\sc bb + cpl} consisting of a simple
blackbody identified with emission from the neutron star, plus a cut-off power law
to represent Comptonization provided good fits to all spectra. 
It may not appear that a cut-off power law is a sufficiently complex
form to describe Comptonization. The forms of {\it Xspec} models
such as {\sc thcompds}, {\sc thcompfe} and {\sc comptt} 
differ from a cut-off power law in the shape of the knee.
We have carried out simulations that show that unless data of the highest quality
are available, i.e. with very small Poisson errors, which are broadband, i.e.
extending well above the knee, there is little error in using a cut-off
power law. The knee occurs between 10 -- 100 keV in all but the brightest sources,
so broadband data normally means in the energy range 1 -- 100 keV.
For such high quality broadband data, use of the
{\sc cpl} model would give somewhat inaccurate values of the power law index, and the
thermal Comptonization model suggested by the present work should be used.
In our previous work, we have not had available data of sufficient quality
that using a cut-off power law has caused significant errors.

In Sect. 4, we addressed the correct form for Comptonization models, 
and showed that we expect the spectrum to continue to rise below 1 keV.
It was recently claimed that the Birmingham model
was not a correct description of Comptonization (Done et al. 2002), since we
do not include the marked turnover at low energies inherent in the Eastern model
(Fig. 3, right panel). If we neglect the turn-over, we overestimate
the Comptonized emission and so underestimate blackbody emission. This might
explain the unphysically small values of inner disc radius, in many cases less than 0.5 km,
we obtained in applying the Eastern model (in the form disc blackbody plus cut-off power law)
in a survey of LMXB with {\it ASCA}. The present work demonstrates that with an extended ADC, 
there will not be a low energy cut-off in the spectrum and so the above argument
that the blackbody term is underestimated does not appear valid.

%

\section*{Acknowledgments}
We thank Dr. Piotr \.Zycki for providing the {\sc thcompds} model.
This work was supported in part by the Polish KBN grants
PBZ-KBN-054/P03/2001 and KBN-2P03D-01525/2003.

\end{document}